\documentclass[prd,superscriptaddress,amsmath,notitlepage,onecolumn,12pt]{revtex4-1}
\usepackage{graphicx}
\usepackage{float}
\usepackage{epstopdf,cancel}
\usepackage{epsf,latexsym,bbm,euscript}
\usepackage{amssymb,amsmath}
\usepackage{mathtools} 
\usepackage{times,graphics}
\usepackage{soul,xcolor}
\usepackage{mathtools}
\usepackage[normalem]{ulem}
\newcommand{\D}{\mathrm{d}}

\def\6{{\langle}}
\def\9{{\rangle}}

\newcommand{\be}{\begin{equation}}
\newcommand{\ee}{\end{equation}}
\newcommand{\ba}{\begin{eqnarray}}
\newcommand{\ea}{\end{eqnarray}}

\usepackage{xtab,afterpage}
\usepackage{hyperref}
\usepackage{setspace,calc}

\begin{document}

\title{Signatures of discretization in quantum black hole spectra}
 
\author{Joshua Foo\footnote{Corresponding Author}}
\email{joshua.foo@uqconnect.edu.au}
\affiliation{Centre for Quantum Computation \& Communication Technology, School of Mathematics \& Physics, The University of Queensland, St.~Lucia, Queensland, 4072, Australia}

\author{Robert B.\ Mann}
\email{rbmann@uwaterloo.ca}
\affiliation{Department of Physics and Astronomy, University of Waterloo, Waterloo, Ontario N2L 3G1, Canada}
\affiliation{Perimeter Institute for Theoretical Physics, Waterloo, Ontario N2L 6B9, Canada}

\author{Magdalena Zych}
\email{magdalena.zych@fysik.su.se}
\affiliation{Department of Physics, Stockholm University, AlbaNova University Center, SE-106 91 Stockholm, Sweden}
\affiliation{Centre for Engineered Quantum Systems, School of Mathematics and Physics, The University of Queensland, St. Lucia, Queensland, 4072, Australia}

\begin{abstract}
\vspace*{1mm}

The quantum superposition principle states that quantum-mechanical systems such as atoms can be placed in a superposition of mass-energy eigenstates. Inspired by this idea and the seminal conjecture of Bekenstein, who proposed that black holes in quantum gravity must possess a discrete mass eigenspectrum, here we analyze the effects produced by a black hole in a superposition of masses. Analogous to using the electromagnetic field to probe atoms, we consider a quantum scalar field on the spacetime background sourced by the black hole mass superposition. From the resulting spectra, as measured by a hypothetical two-level system interacting with the field, we infer signatures of discretization of the black hole mass in support of Bekenstein's conjecture. 

\end{abstract}

\maketitle

\vspace{30mm}

\textit{Essay written for the Gravity Research Foundation 2023 Awards for Essays on Gravitation.}

\vspace{15mm}

\text{\hfill \hspace*{104mm} Submitted on March 29, 2023}

\vspace{5mm}

\newpage

Black holes have long captivated physicists from a diverse array of backgrounds, from cosmology and astroparticle physics to quantum field theory and general relativity. Because of the extreme gravitational environments they generate, they are considered primary candidates for studying regimes in which quantum-gravitational effects are present \cite{gibbons1993euclidean,wald1984black,nicolinidoi:10.1142/S0217751X09043353}. Indeed, the discoveries of Hawking radiation \cite{hawking1974black} and black hole evaporation \cite{hawkingcmp/1103899181} gave rise to the well-known information paradox \cite{hawkingPhysRevD.14.2460,Giddings:1995gd,Mathur_2009,Harvey:1992xk} and an entire field seeking its resolution, which aptly illustrates the existing conflicts between quantum theory and general relativity. 

Bekenstein \cite{bekensteinPhysRevD.7.2333,bekenstein2020quantum,reggePhysRev.108.1063} was among the first to recognise that a complete theory of quantum gravity must account for the treatment of black holes as quantum objects. One of his key insights was in recognising that the black hole horizon area, and hence its mass, is an adiabatic invariant, $I$, with an associated discrete quantization \cite{hodPhysRevLett.81.4293} of evenly spaced energy levels \cite{BEKENSTEIN19957}. That is, 
\begin{align}\label{eq1}
    I &= \int\D A \sim \int \: \D M \: f(d,M) \equiv n \hslash \omega ,
\end{align}
where $A$ is the horizon area, $f(d,M)$ is some function of the mass $M$ and spacetime dimension $d$, $\omega$ is the transition frequency of the black hole, and $n \in \mathbb{Z}$. The first equality of Eq.\ (\ref{eq1}) is Bekenstein's identification of the horizon area as an adiabatic invariant; this gives rise, via the first law of black hole thermodynamics, to the middle relationship, while the last equality describes the equally-spaced discretization of the horizon area and mass.

Following Bekenstein's analogical treatment of black holes as atomic systems motivates us to consider them in quantum superpositions of mass-energy eigenstates. A mass-superposed black hole is an example of a non-classical spacetime; since the masses individually characterize a unique solution to Einstein's field equations in classical general relativity, the resulting scenario can be described as an entangled state where the masses and the associated states of spacetime are jointly superposed \cite{fooPhysRevLett.129.181301,fooPhysRevD.107.045014,Foo_2021desitter,foodoi:10.1142/S0218271822420160}. Understanding the effects that arise in such spacetime superpositions is an important stepping stone towards developing a complete description of a quantum spacetime. Formal approaches for quantizing general relativity date back to the Wheeler-deWitt equation \cite{Wheeler1968SUPERSPACEAT,dewittPhysRev.160.1113}, which governed a wave functional of the metric and offered a conceptual scheme for dealing with superpositions of spacetime geometries \cite{rovelli2015covariant,rovelli2008loop}. Building upon this work, canonical quantization techniques were applied to the metric variables \cite{kuchaPhysRevD.50.3961,THIEMANN1993211} leading to the development of loop quantization \cite{ashtekarPhysRevLett.57.2244,ashtekarPhysRevLett.80.904}, which in turn has yielded solutions including black holes in a superposition of masses \cite{rovelli2004quantum,KASTRUP1994665,Campiglia_2007,Gambini_2014,gambiniPhysRevLett.110.211301,Demers_1996,Kiefer_2013}.

In this essay, we outline an \textit{operational} approach for studying the phenomenology of black hole mass superpositions. By operational, we mean that our approach is grounded in effects that can in-principle be measured by physical systems such as rods, clocks, and other measurement devices. In particular, we consider a two-level system that interacts with
a scalar quantum field on a background sourced by a (2+1)-dimensional black hole in a superposition of masses. We show how the dynamics of this system can probe the black hole's ``spectrum,'' in analogy with spectroscopic experiments that probe atomic systems via the electromagnetic field. As we will demonstrate, the spectrum measured by the two-level system elicits signatures of discretization, in accordance with Bekenstein's original conjecture for quantum black holes. 

Let us now construct the scenario described above. We focus on the (2+1)-dimensional Banados-Teitelboim-Zanelli (BTZ) black hole of mass $M$, whose line element is given by \cite{banadosPhysRevLett.69.1849,Carlip:1995qv,carlip2003quantum}
\begin{align}\label{eq2}
    \D s^2 &= - \left( \frac{r^2}{l^2} - M \right) \D t^2 + \left( \frac{r^2}{l^2} - M \right)^{-1} \D r^2 + r^2 \D \phi^2 ,
\end{align}
where $-l^2= \Lambda $ is the AdS length (corresponding to a negative cosmological constant $\Lambda$). Let us consider the two-level probe system situated at a fixed coordinate distance $r = R_D$ from the origin and prepared in its ground state, $| g \rangle$. The scalar field is assumed to be initially in the AdS vacuum, $| 0 \rangle$, while the black hole is initially in a superposition of two different masses $M_A, M_B$. Note that Eq.~\eqref{eq2} applies for any value of the mass $M$, and thus we can use the same coordinates $t$, $r$, $\phi$ for both masses. Within quantum theory, we can denote the mass eigenstates of the black hole as $| M_A \rangle$, $| M_B \rangle$, and we assume these states are orthogonal (as different black hole masses correspond to classically distinguishable situations). The state of the entire system at some initial coordinate time $t = t_i$ is thus given by,
\begin{align}
    | \psi(t_i) \rangle &= \frac{1}{\sqrt{2}} ( | M_A \rangle + | M_B \rangle ) \otimes | 0 \rangle \otimes | g \rangle . 
\end{align}
The interaction between the black hole, quantum field, and the probe can be described with a Hamiltonian of the form \cite{fooPhysRevLett.129.181301} 
\begin{align}\label{Hint}
    \hat{H}_\mathrm{int} &= \hat{H}_A \otimes | M_A \rangle \langle M_A | + \hat{H}_B \otimes | M_B \rangle\langle M_B | ,
\end{align}
where  $\hat{H}_{D}$ $(D = A,B)$ is a Hamiltonian describing the probe system and the field in the spacetime generated by the black hole of mass $M_D$, in particular with the line element in Eq.~\eqref{eq2} with $M=M_D$. Specifically, we consider a linear coupling between the probe and the field $\hat{H}_{D} \propto \hat{\phi}(\mathbf{x}_{D}) \otimes ( | e \rangle \langle g | e^{i\Omega\tau} + \mathrm{H.c.})$ where $\Omega$ is the internal energy gap of the two-level probe and $\hat{\phi}(\mathbf{x}_{D})$ is the field operator pulled back to the worldline $\mathbf{x}_{D}$ in the spacetime generated by mass $M_D$. The Hamiltonian Eq.~\eqref{Hint} describes how for each of the black hole masses, the field and probe interact as in a fixed spacetime characterized by that mass. 

Time-evolution of the system in the interaction picture up to some final time $t = t_f$ takes the form, 
\begin{align}\label{psifin}
    | \psi(t_f) \rangle &= e^{-i\hat{H}_{0,S}t_f} \hat{U}(t_i,t_f) e^{i\hat{H}_{0,S}t_i} | \psi(t_i) \rangle 
\end{align}
where $\hat{H}_{0,S}$ is the free Hamiltonian of the system, including the free evolution of the mass-energy eigenstates of the black hole.  Here, the coordinate time $t$ used to parametrize the evolution of all systems can be defined as the proper time of a clock at sufficiently large $r$, such that the fractional difference in proper times between the two spacetimes, $(M_B - M_A)l^2/r$, is negligible. This choice assumes that the state of the clock is factorizable, i.e.\ independent of the black hole mass. One could alternatively choose a clock state that is entangled with the mass; in an anticipated theory of quantum gravity, one expects both scenarios to be physically admissible.

Now, recall that we are motivated by the notion of \textit{operationalism}. In the present case, we seek to describe effects that are \textit{in-principle detectable} via measurements of a probe due to its interaction with a field quantized on a spacetime in a superposition of masses. In the absence of such an interaction, one could hypothetically complete a ``double-slit experiment'' with the black hole masses, leading to the usual interference fringes between them. However this would reveal little information about the structure of the non-classical spacetime sourced by the black hole, and the effect of such an environment on physical systems residing within. Again drawing an analogy with spectroscopic experiments used to image atomic wavefunctions, here we propose to ``image'' the spacetime produced by the black hole through its coupling with the field and probe. This is the crux of what makes our approach an operational one. 

To this end, let us calculate the transition probability of the two-level probe from its ground to excited state \ $P ( | g \rangle \to | e \rangle )$. One way to obtain this is to trace out both the field and black hole degrees of freedom. If we apply this to the final state in Eq.~\eqref{psifin}, we obtain
\begin{align}\label{eq5}
    \mathrm{Tr}_{\phi,M} \Big[ | \psi(t_f) \rangle\langle \psi(t_f) | \Big] &= \frac{1}{2} ( P_A + P_B ),
\end{align}
which is a statistical mixture of the transition probabilities $P_A$, $P_B$ for a detector situated in the classical spacetime sourced by a black hole of \textit{either} mass $M_A$ or $M_B$. In this case, the quantum features of the spacetime superposition are washed out. 

A more interesting case for us is 
when the black hole mass is assumed to be measured in a basis consisting of two orthogonal mass eigenstates, e.g.\ $| \pm \rangle = ( | M_A \rangle \pm | M_B \rangle ) / \sqrt{2}$, and tracing out the field. This amounts to completing a Mach-Zehnder-type interferometric protocol on the black hole masses, with the field and probe interacting with the black hole in each branch of the interferometer. A measurement of the black hole mass could be achieved by an interaction of an ancilla with the black hole and later measuring the ancilla in a superposition basis. More specifically, one could prepare two qubits entangled in their internal energies, before sending one into the black hole thereby entangling its mass with the state of other qubit. Projecting this qubit onto a superposition state will in turn project the black hole onto a mass superposition state. The result of such a protocol gives
\begin{align}\label{conditionalprob}
    \mathrm{Tr}_{\phi} \Big[ \langle \pm | \psi(t_f) \rangle\langle \psi(t_f) | \pm \rangle \Big] &= \frac{1}{4} ( P_A + P_B \pm 2 \cos ( \Delta E \Delta t ) L_{AB} ), 
\end{align}
where $\Delta t=t_f-t_i$ is the coordinate time elapsed, $\Delta E = | \sqrt{M_B} - \sqrt{M_A} |$ is the energy gap of the black hole eigenstates, and we refer to $L_{AB}$ as the cross-term, which we discuss in detail below. We note that the above quantity is a conditional transition probability of the probe, conditioned on the outcome $\pm$ of the measurement of the black hole mass in the basis  $| \pm \rangle$.

\begin{figure}[h]
    \centering
    \includegraphics[width=0.6\linewidth]{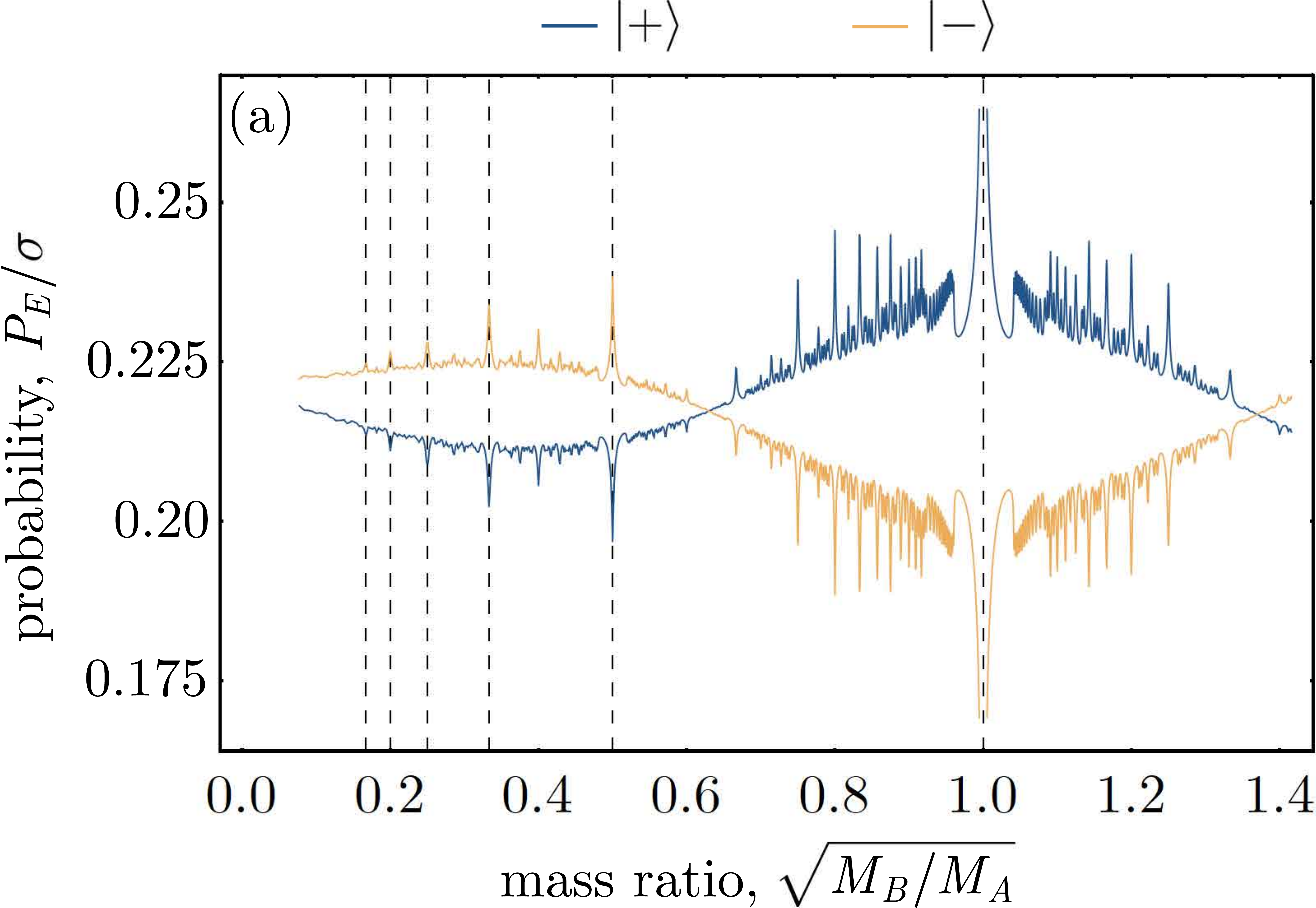}
    \caption{Transition probability of a two-level detector situated outside a black hole in a superposition of masses. The blue and yellow lines correspond to measurements of the black hole in the $| + \rangle$ and $| - \rangle$ state respectively. The transition probability is normalized by $\sigma$, the timescale of the interaction.}
    \label{fig:1}
\end{figure}

Equation \eqref{conditionalprob} shows that the probe transition probability is now sensitive to interference between the different spacetimes in superposition, information of which is contained in the cross-term $L_{AB}$. Notably, this cross-term takes the form \cite{fooPhysRevLett.129.181301}
\begin{align}\label{eq8}
    L_{AB} &\propto \sum_{n,m} \mathrm{Re} \int_0^{t_f/l} \D z \: F\left( z , n \sqrt{M_A}  - m \sqrt{M_B} \right) , \: n , m \in \mathbb{Z} ,
\end{align}
where $F(\cdot , \cdot)$ is a complicated function of its arguments. The dependence of $L_{AB}$ (and hence the transition probability) on $n \sqrt{M_A} - m \sqrt{M_B}$ is suggestive of nontrivial behaviour 
when $\sqrt{M}_A$, $\sqrt{M}_B$ are integer-valued. Indeed, this is what we find. Figure.\ \ref{fig:1} displays the transition probability  as a function of the ratio $\sqrt{M_B/M_A}$, where resonances are clearly seen at rational values of $\sqrt{M_B/M_A}$; for example $\sqrt{M_B/M_A} = 1/4, 1/3$ and so on.

Remarkably, the location of these resonances corresponds with values arising from Bekenstein's conjecture that the mass spectrum of a black hole is discrete. In particular, for the case of the BTZ black hole \cite{Kwon_2010}, 
\begin{align}
    I = l \sqrt{M} = n \hslash , \: n \in \mathbb{Z}, 
\end{align}
implying, that for any two black hole masses $M_A$ and $M_B$, $\sqrt{M_B/M_A} \in \mathbb{Q}$. Importantly, our construction does not make any assumptions about the masses $M_A, M_B$, but the form of the resulting transition probability, Eq.\ (\ref{eq8}), gives rise to a spectrum--as detected by a two-level system coupled to the field--that encodes a novel signature of mass discretization. We therefore suggest that the result above constitutes an independent verification of--indeed an extension to (for superpositions of 
masses)--Bekenstein's conjecture.  More significantly, it is also a demonstration of how operational techniques can be used to reveal genuinely quantum-gravitational properties of a black hole.

The scheme presented here opens a path to understanding the effects of non-classical spacetimes on important notions such as causal structure or reference frames in quantum gravity. Concerning the former, our approach provides techniques for studying quantum communication protocols in such spacetimes, and the possibility of violating classical constraints on signalling (for example, considering multiple probes situated outside a mass-superposed black hole and signalling between them through the field). Concerning the latter, the mass-quantised black hole considered here naturally gives rise to two distinct sets of coordinates--correlated as well as uncorrelated with the mass. An open question remains as to how one appropriately defines the resulting transformations between them, and how they act on the systems in question (i.e.~the field and the probe system).

Our results also motivate the development of experimental setups that can simulate the operational effects of spacetime in quantum superposition, utilizing for example, advances in optomechanical technologies. The line element of Eq.\ (\ref{eq2}) is obtained by imposing a periodic boundary condition on AdS spacetime; in this sense, the mass-superposed BTZ black hole is an example of a quantum superposition of non-trivial topologies. Such topologies have been realized in superfluid Helium condensates on toroidal cavities. The fundamental mode providing the analog metric for phonons in the condensate therein can interact with incident photons, thus suggesting an opportunity to study new analog quantum-gravitational systems, in which a quantum field evolves in the presence of quantum-controlled backgrounds
\cite{bowen2015quantum}.



\bibliography{References.bib}


\end{document}